\documentstyle[epsf,12pt]{article}
\input epsf
\newlength{\dinwidth}
\newlength{\dinmargin}
\begin{document}
\newcommand{\Gevsq}     {\mbox{${\rm GeV}^2$}}
\newcommand{\qsd}       {\mbox{${Q^2}$}}
\newcommand{\rhod}      {\mbox{${\rho^0}$}}
\newcommand{\yjb}       {\mbox{${y_{_{JB}}}$}}
\newcommand{\xda}       {\mbox{$x_{_{DA}}$}}
\newcommand{\qda}       {\mbox{$Q^{2}_{_{DA}}$}}
\newcommand{\gsubh}     {\mbox{$\gamma_{_{H}}$}}
\newcommand{\thee}      {\mbox{$\theta_{e}$}}
\newcommand{\mrsdm}     {\mbox{MRSD$_-^{\prime}$\ }}
\newcommand{\mrsdz}     {\mbox{MRSD$_0^{\prime}$\ }}
\newcommand{\sleq} {\raisebox{-.6ex}{${\textstyle\stackrel{<}{\sim}}$}}
\newcommand{\sgeq} {\raisebox{-.6ex}{${\textstyle\stackrel{>}{\sim}}$}}
\renewcommand{\thefootnote}{\arabic{footnote}}
\def\pom{{\cal P}}

\title{ \vspace{2cm}
{\bf Exclusive $\rhod$ production in deep inelastic
electron-proton scattering at HERA}\\
\author{\rm ZEUS Collaboration}  }
\date{}
\maketitle
\vspace{5 cm}
\begin{abstract}
\par
The exclusive production of $\rhod$ mesons in deep inelastic
electron-proton scattering has been studied using the ZEUS detector.
Cross sections have been measured in the range $7 < Q^2 < 25$ GeV$^2$ for
$\gamma^*p$ centre of mass (c.m.) energies from 40 to 130 GeV. The
$\gamma^*p \rightarrow \rhod p$ cross section exhibits a $Q^{-(4.2 \pm
0.8 ^{+1.4}_{-0.5})}$ dependence and both longitudinally and transversely
polarised $\rhod$'s are observed. The $\gamma^*p \rightarrow \rhod p$
cross section rises strongly with increasing c.m. energy, when compared
with NMC data at lower energy, which cannot be explained by production
through soft pomeron exchange. The data are compared with perturbative
QCD calculations where the rise in the cross section reflects the
increase in the gluon density at low $x$. the gluon density at low $x$.
\end{abstract}
\vspace{-19cm}
\begin{flushleft}
\tt DESY 95-133 \\
July 1995 \\
\end{flushleft}

\setcounter{page}{0}
\thispagestyle{empty}
\newpage

\def\3{\ss}
\footnotesize
\renewcommand{\thepage}{\Roman{page}}
\begin{center}
\begin{large}
The ZEUS Collaboration
\end{large}
\end{center}
M.~Derrick, D.~Krakauer, S.~Magill, D.~Mikunas, B.~Musgrave,
J.~Repond, R.~Stanek, R.L.~Talaga, H.~Zhang \\
{\it Argonne National Laboratory, Argonne, IL, USA}~$^{p}$\\[6pt]
R.~Ayad$^1$, G.~Bari, M.~Basile,
L.~Bellagamba, D.~Boscherini, A.~Bruni, G.~Bruni, P.~Bruni, G.~Cara
Romeo, G.~Castellini$^{2}$, M.~Chiarini,
L.~Cifarelli$^{3}$, F.~Cindolo, A.~Contin, M.~Corradi,
I.~Gialas$^{4}$,
P.~Giusti, G.~Iacobucci, G.~Laurenti, G.~Levi, A.~Margotti,
T.~Massam, R.~Nania, C.~Nemoz, \\
F.~Palmonari, A.~Polini, G.~Sartorelli, R.~Timellini, Y.~Zamora
Garcia$^{1}$,
A.~Zichichi \\
{\it University and INFN Bologna, Bologna, Italy}~$^{f}$
% ...
\\[6pt]
\clearpage
\noindent
\hspace*{1mm}
$^{ 1}$ supported by Worldlab, Lausanne, Switzerland \\
\hspace*{1mm}
$^{ 2}$ also at IROE Florence, Italy  \\
\hspace*{1mm}
$^{ 3}$ now at Univ. of Salerno and INFN Napoli, Italy  \\
\hspace*{1mm}
$^{ 4}$ supported by EU HCM contract ERB-CHRX-CT93-0376 \\
\hspace*{1mm}
% ...
$^{18}$ now with OPAL Collaboration, Faculty of Physics at Univ. of
        Freiburg \\
% ...

\begin{tabular}{lp{15cm}}
$^{a}$ &supported by the Natural Sciences and Engineering Research
         Council of Canada (NSERC) \\
$^{b}$ &supported by the FCAR of Qu\'ebec, Canada\\
$^{c}$ &supported by the German Federal Ministry for Research and
         Technology (BMFT)\\
$^{d}$ &supported by the MINERVA Gesellschaft f\"ur Forschung GmbH,
         and by the Israel Academy of Science \\
$^{e}$ &supported by the German Israeli Foundation, and
         by the Israel Academy of Science \\
$^{f}$ &supported by the Italian National Institute for Nuclear Physics
         (INFN) \\
$^{g}$ &supported by the Japanese Ministry of Education, Science and
         Culture (the Monbusho)
         and its grants for Scientific Research\\
$^{h}$ &supported by the Korean Ministry of Education and Korea Science
         and Engineering Foundation \\
$^{i}$ &supported by the Netherlands Foundation for Research on Matter
         (FOM)\\
$^{j}$ &supported by the Polish State Committee for Scientific Research
         (grant No. SPB/P3/202/93) and the Foundation for Polish-
         German Collaboration (proj. No. 506/92) \\
$^{k}$ &supported by the Polish State Committee for Scientific
         Research (grant No. PB 861/2/91 and No. 2 2372 9102,
         grant No. PB 2 2376 9102 and No. PB 2 0092 9101) \\
$^{l}$ &partially supported by the German Federal Ministry for
         Research and Technology (BMFT) \\
$^{m}$ &supported by the German Federal Ministry for Research and
         Technology (BMFT), the Volkswagen Foundation, and the Deutsche
         Forschungsgemeinschaft \\
$^{n}$ &supported by the Spanish Ministry of Education and Science
         through funds provided by CICYT \\
$^{o}$ &supported by the Particle Physics and Astronomy Research
        Council \\
$^{p}$ &supported by the US Department of Energy \\
$^{q}$ &supported by the US National Science Foundation
\end{tabular}

\newpage
\pagenumbering{arabic}
\setcounter{page}{1}
\normalsize

\section{\bf Introduction}

\par
With the high energy electron-proton collider HERA, it has
become possible to study deep inelastic scattering (DIS) processes at
large values of $Q^2$, the negative of the four-momentum transfer squared
of the exchanged
virtual photon, and large values of $W$, the virtual photon-proton centre
of mass (c.m.) energy.  The exclusive production of vector mesons in DIS
is of particular interest.
While numerous data exist at low $Q^2$  [1-5] on the exclusive reaction
\begin{equation}
   e~(or~\mu)~+~p \rightarrow e~(or~\mu)~  +~\rhod~+p,
\end{equation}
\noindent
only two experiments have reported DIS
measurements for $Q^2 > $ 5 GeV$^2$ \cite{emc,nmc}.
These latter measurements have been restricted to $W<20$ GeV.

Previous studies of exclusive leptoproduction ($\gamma^{*}N \rightarrow
\rhod N$) and real photoproduction ($\gamma N \rightarrow \rhod N$)
off a nucleon $N$ have shown
that for $\rhod$ production at low $Q^2$ (typically $<$ 2 GeV$^2$):
\begin{itemize}
\item the $Q^2$ dependence of the cross section can be described by the
    Vector Dominance Model (VDM) \cite{bauer} in which it is assumed that
    the photon fluctuates into a $\rhod$ meson yielding:
\begin{equation}
  \frac{d\sigma(Q^2)}{dt} = \frac{d\sigma(0)}{dt} \left (\frac{M_{\rho}^2}
  {M_{\rho}^2 + Q^2} \right)^2 \left(   {1+\epsilon \xi^2 \frac{Q^2}
  {M_{\rho}^2}} \right) e^{bt},
\end{equation}
   where $\xi$ is the ratio of the longitudinal to transverse forward
   amplitudes, $\epsilon$ is the relative longitudinal
   polarisation of the virtual
   photon and $M_{\rho}$ is the $\rhod$ mass; the distribution of $t$,
   the square of the four-momentum transfer between the photon and the
   $\rhod$, is described by a single exponential dependence, in the range
   from $t$ = 0 to $t$ = $-0.5$ GeV$^2$, with a
   slope parameter, $b\approx 7-12$ GeV$^{-2}$;
\item in real photoproduction ($Q^2$ = 0),
 the process is `quasi-elastic' and the helicity
of the $\rhod$ is similar to that of the incident photon, i.e. s-channel
helicity is largely conserved (SCHC)~\cite{sbt}.
The $\rhod$ decay distribution exhibits an approximate $\sin^2
\theta_{h}$ dependence, where $\theta_h$ is the polar angle of the $\pi^+$ in
the $\rhod$ c.m. system and the quantisation axis is the
$\rhod$ direction in the $\gamma p$ c.m. system;
\item the  real photoproduction $\rhod$ total cross section increases
slowly as a function of $W$ for $W> 15$ GeV~\cite{rhot,spprho}. This is
expected for an elastic reaction dominated by the
exchange of a `soft' pomeron with an intercept of the
 Regge trajectory of $\alpha(0)
= 1+\epsilon^{\prime}$ = 1.08. The intercept is determined from fits~\cite{dl1}
to hadron-hadron total cross sections: for a $\gamma p$ total cross
section $\sigma_{tot} \sim W^{2(\alpha(0)-1)} \sim  W^{2\epsilon^{\prime}}$,
the optical theorem yields
$ \left. \frac{d\sigma_{el}}{dt}\right|_{t=0} \sim  W^{4\epsilon^{\prime}}
\sim W^{0.32}$.
\end{itemize}

At larger $Q^2$ ($2 < Q^2 < 25$ GeV$^2$), leptoproduction results from
the EMC \cite{emc} and NMC \cite{nmc} experiments indicate that:
\begin{itemize}
\item  the $\gamma^{*}p \rightarrow \rhod p$ cross section
is consistent with a 1/$Q^4$ behaviour;
\item at larger $Q^2$ ($>6$ GeV$^2$), the distribution of the square of
the transverse momentum of the $\rhod$ with respect to the virtual photon
 ($p_T^2$) is exponentially falling with a slope of $b=4.6 \pm
0.8$~GeV$^{-2}$~\cite{nmc},
about a factor of two smaller than that of the photoproduction elastic
process;
\item as $Q^2$ increases,
the fraction of zero-helicity, longitudinally polarised
$\rhod$s increases above 50\%; assuming SCHC
\cite{bauer}, this implies that the longitudinal virtual photon cross section
dominates;
\item for $2 < W < 4$ GeV the cross section falls with $W$ at small
$Q^2 < 4$ GeV$^2$~\cite{lame}. No significant  dependence on $W$ is observed
for $9<W<19$ GeV \cite{nmc}.
\end{itemize}

The reaction $\gamma^{*}p \rightarrow \rhod p$ has also been the focus of
theoretical investigations. Early studies based on VDM are
described elsewhere~\cite{bauer}. A study of diffractive leptoproduction by
Donnachie and Landshoff based on a zeroth order perturbative QCD
(pQCD) calculation for pomeron exchange at small values of Bjorken $x \sim
Q^2/W^2$ \cite{dl1}
reproduced many of the features seen experimentally, including the $Q^2$
dependence of the data at
low $W$.  In more recent presentations, the pomeron is treated as a
non-perturbative two-gluon exchange~\cite{dl2}.
This approach has also been studied by
Cudell~\cite{cudell}.  Calculations in pQCD have been performed in the leading
logarithm approximation for $J/\psi$ electroproduction by
Ryskin~\cite{ryskin}.
Ginzburg et al.~\cite{ginz} and Nemchik et al.~\cite{nnz}
have also performed a calculation in pQCD for
vector meson production.
These more recent calculations predict a $Q^{-6}$
dependence of the longitudinal cross section at high $Q^2$, in contrast to
the $Q^{-4}$ VDM expectation. Brodsky et al.~\cite{brod} have studied the
forward scattering cross section for this reaction by applying pQCD
in the double leading logarithm approximation (DLLA).
They predict that at high $Q^2$ the
vector mesons should be produced dominantly by longitudinally polarised
virtual photons with a
dependence for the longitudinal cross section:
\begin{equation}
\label{brodsky}
 \left. \frac{d\sigma _L}{dt}\right|_{t=0}(\gamma^*N \rightarrow \rhod N) =
\frac{A}{Q^6} \alpha_s^2(Q^2) \cdot \left| \left[ 1 + i(\pi/2)
(\frac{d}{d~ln~x})
\right] xg(x,Q^2) \right|^2,
\end{equation}
where $A$ is a constant, which can be calculated,
and $xg(x, Q^2)$ is the momentum density of the gluon in the
proton.
Using the $x$-dependence of $xg(x,Q^2)$ measured at HERA and the relation
$W^2 \sim Q^2/x$ for small $x$, at fixed $Q^2$ one expects
that $ \left. \frac{d\sigma_{el}}{dt}\right|_{t=0}
 \sim W^{1.4}$ \cite{zgluon}, in contrast to the
$W^{0.32}$ dependence expected for the `soft' pomeron.

This letter presents a measurement with the ZEUS detector
of the exclusive
cross section for $\rhod$ mesons produced at large $Q^2$
by the virtual  photoproduction
process $\gamma^{*}p \rightarrow \rhod p$ at HERA.
The data come from neutral current, deep inelastic
electron-proton scattering in
the $Q^2$ range of 7 - 25 GeV$^2$, similar
to that of the earlier fixed target experiments \cite{emc,nmc};
however, they cover a lower $x$
region ($4\cdot10^{-4} < x < 1\cdot10^{-2}$) and, consequently, a higher
$W$ region (40-130 GeV).

\section{ Experimental conditions}
The experiment was performed at the electron-proton collider HERA
using the ZEUS detector. During 1993 HERA operated with bunches of
electrons of energy $E_e=26.7$ GeV colliding
with bunches of protons of energy $E_p=820$ GeV, with a time interval
between bunch
crossings of 96 ns. For this data-taking period
84 bunches were filled for each beam (paired bunches)
and in addition 10 electron and 6 proton bunches were left unpaired
for background studies. The electron and proton beam
currents were typically 10 mA.  The $ep$ c.m. energy is $\sqrt{s}$ =
296 GeV and
the integrated luminosity was 0.55 pb$^{-1}$.

ZEUS is a multipurpose magnetic detector whose 1993 configuration
has been described elsewhere \cite{zf2}. This brief description concentrates
on those parts of the detector relevant to the present analysis.

Charged particles
are tracked by the inner tracking detectors which operate in a
magnetic field of 1.43 T.
Immediately surrounding the beampipe is the vertex detector
(VXD)~\cite{vxd} which
consists of 120 radial cells, each with 12 sense wires.
Surrounding the VXD is the central tracking detector (CTD)~\cite{ctd}
which consists of 72 cylindrical
drift chamber layers,  organised into 9 `superlayers'.
In events with charged
particle tracks, using the combined data from both chambers,
resolutions of $0.4$ cm in $Z$ and $0.1$ cm in radius
in the $XY$ plane\footnote{The ZEUS coordinate system
is defined as right-handed with the $Z$ axis pointing in the
proton beam direction, hereafter referred to as forward, and the $X$
axis horizontal, pointing towards the centre of HERA.}
are obtained for the primary vertex reconstruction.
These detectors provide a momentum resolution given by $\sigma_{p_T}/p_T
= \sqrt{(0.005p_T)^2 + (0.016)^2}$ (with $p_T$ in GeV).

The superconducting
solenoid is surrounded by a high resolution  uranium/scintillator
calor- imeter which is divided into three parts:
forward (FCAL), barrel (BCAL), and rear (RCAL) covering the angular region
$\rm 2.2^{o}<\theta<176.5^{o}$,
where $\rm \theta=0^{o}$ is defined as the proton beam direction.
Holes of $20\times 20$ cm$^2$ in the centre of FCAL and
RCAL are required to accommodate the HERA beam pipe.
The calorimeter parts are
subdivided into towers which in turn are subdivided longitudinally
into electromagnetic
(EMC) and hadronic (HAC) sections. The sections are subdivided into cells,
each of which is viewed by two photomultiplier tubes which provide the
energy and the time of the energy deposit with a resolution of better
than 1 ns.
An additional hadron-electron separator (RHES)\cite{hes},
located at the electromagnetic shower maximum in
the RCAL and
consisting of a layer of $3 \times 3$ cm$^2$
 silicon diodes, was used to provide more accurate position information for
electrons scattered  at low angles than was available from the
calorimeter alone.


The luminosity is measured from the rate observed in the luminosity
photon detector of hard bremsstrahlung photons
from the Bethe-Heitler process $ep \rightarrow e p\gamma$.
The luminosity detector consists of photon and electron
lead-scintillator calorimeters~\cite{lumi}.
Bremsstrahlung photons
emerging from the electron-proton interaction
point at angles below 0.5 mrad with respect to the
electron beam axis hit the photon calorimeter placed 107 m along the
electron beam line. Electrons emitted at scattering angles
less than 5 mrad and with energies
$0.2 E_e <E_e^{\prime} < 0.9 E_e$ are deflected by beam magnets and
hit the electron calorimeter placed 35~m from the interaction point.

The data were collected with a three-level-trigger. The
first-level-trigger (FLT)
for DIS events required a logical OR of three conditions on sums of
energy in the EMC calorimeter cells.
Details are given elsewhere \cite{zf2,wsmith}. For events  with
the scattered electron detected in the calorimeter, the FLT was essentially
independent of the DIS hadronic final state. The FLT acceptance was greater
than 97\% for $Q^2 > 7~\mbox{\rm GeV}^2 $.
The second-level-trigger used information from a subset of detector
components to
reject proton beam-gas events, thereby reducing the FLT  DIS triggers
by an order of magnitude, but without loss of DIS events.

The third-level-trigger (TLT) had available the full event information
on which to apply physics-based filters. The TLT
applied stricter cuts on the event times and also rejected beam-halo
muons and cosmic ray muons. Events remaining after the above veto cuts
were selected for output by the TLT if $\delta\equiv\Sigma_i
E_i(1-\cos\theta_i) > 20~\mbox{GeV} - 2E_{\gamma}$,
                      where $E_i,~\theta_i$ are the energy and polar angle
(with respect to the nominal beam interaction point) of the geometric
centre of a calorimeter cell and
$E_{\gamma}$ is the energy measured in the photon calorimeter of the
luminosity monitor.
For fully contained events $\delta\sim 2E_e=53.4$ GeV.
For events from photoproduction,  the scattered electrons remain in the rear
beam pipe and $\delta$ peaks at low values.

\section{Kinematics of exclusive $\rhod$ production}

The kinematic variables used to describe $\rhod$
production in the reaction:
\begin{equation}
      e~+~p \rightarrow e~+~\rhod~+X,
\end{equation}
\noindent
where $X$ represents either a proton or a diffractively dissociated
proton remnant of mass $M_X$,
are the following:  the negative of the squared four-momentum
transfer carried by the virtual photon\footnote{In the $Q^2$ range
covered by this data sample, effects due to $Z^0$ exchange can be neglected.}
   $ Q^2=-q^2=-(k~-~k')^2$, where $k$ ($k^{\prime}$) is the
four-momentum of the
incident (scattered) electron; the Bjorken variable $x =\frac{Q^2}{2P\cdot q}$,
where $P$ is the four-momentum of the incident proton;
the variable which describes the energy transfer to the hadronic
final state  $   y =\frac{q\cdot P}{k\cdot P}$;
the c.m. energy, $\sqrt{s}$, of the $ep$
system, where $ s = (k+P)^2$; $W$, the c.m. energy  of the $\gamma^*p$
system:
   $ W^2=(q+P)^2=\frac{Q^2(1-x)}{x}+M_p^2 \approx ys$,
where $M_p$ is the proton mass; and $ t^\prime=|t~-~t_{min}|$,
where $t$ is the four-momentum transfer squared, $t$ = $(q - v)^2 =
(P-P^{\prime})^2$, from the photon
 to the $\rhod$ (with four-momentum $v$),
$t_{min}$ is the minimum kinematically allowed value of $t$ and $P^{\prime}$ is
the four-momentum of the outgoing proton.
The squared transverse momentum $p_T^2$ of the
$\rhod$ with respect to the photon direction is
a good approximation to $t^\prime$  since $t^\prime$ is,
in general, small ($<< 1$~GeV$^2$). For the present
data\footnote{The $-0.08$ GeV$^2$ value
corresponds to $M_X$ = 8 GeV and $Q^2$ = 25 GeV$^2$.}
$t_{min}$ ranges from $-0.0006$ to $-0.08$ GeV$^2$.

In this analysis, the $\rhod$ was observed in the decay $\rhod \rightarrow
\pi^+\pi^-$.  The momentum vector of the $\rhod$ was reconstructed from the
pion momentum vectors determined with the tracking system.
The production angles ($\theta_\rho$ and $\phi_\rho$)
and momentum ($p_{\rho}$) of
the $\rhod$ and the angles of the scattered electron
 (${\theta_e}^\prime$ and ${\phi_e}^\prime$), as determined with
 RCAL and RHES,
were used to reconstruct the kinematic variables $x, Q^2$, etc.
The energy of the scattered electron was determined
from the relation:
\begin{equation}
    E_e^c = \frac{(s+M_{\pi\pi}^2-M_X^2)/2-(E_e+E_p)(E_{\rho}
    -|p_{\rho}|cos\theta_{\rho})}
    {(E_e+E_p)(1-\beta cos\theta_e^\prime)-(E_{\rho}-|p_{\rho}|
    cos\theta_{e \rho})},
\end{equation}
where $E_{\rho}$ is the energy of the $\pi\pi$ pair,
$\theta_{e\rho}$ is the angle between the $\pi\pi$ three vector and the
scattered electron and $\beta$ = $(E_p-E_e)/(E_p+E_e)$.
For the case of  reaction (1), $M_X=M_p$ and $E_e^c$ is a good estimator
of the energy of the scattered electron, $E_e^{\prime}$;
for events in which the proton
diffractively dissociates into the system $X$, $M_X > M_p$ and $E_e^c$
is only slightly different from $E_e^{\prime}$.
The above expression simplifies to
\begin{equation}
   E_e^c \approx [2E_e - (E_{\rho} -|p_{\rho}|cos\theta_{\rho})]/
    (1 - cos\theta_e^\prime)
\end{equation}
when $M_X=M_p$ and the transverse momentum
of the proton is negligible compared to its longitudinal component.
This last relation provides an accurate way to calculate the kinematic
variables and is a simple expression used
to evaluate the radiative corrections for this process.
The variable $y$ is calculated from the expression
$y$ = $ (E_{\rho} -|p_{\rho}|cos\theta_{\rho})/2E_e$.
The calculation of $p_T^2$ also uses the $\rhod$ and electron momenta:
$p_T^2$ = $(p_{ex}+p_{\rho x})^2 + (p_{ey}+p_{\rho y})^2$.

\section{\bf Monte Carlo simulations}

The reaction $ep \rightarrow e\rhod p$ was modelled using two different
Monte Carlo generators. The first, DIPSI \cite{arneo},
describes elastic $\rhod$
production in terms of pomeron exchange with the pomeron treated as
a colourless two-gluon system~\cite{ryskin}. The model
assumes that the exchanged photon fluctuates into a quark-antiquark pair
which then interacts with the two-gluon system.
The cross section is proportional to the square of the gluon
momentum density in the proton. Samples of
$\phi$ and $\omega$ events were generated in a similar way.

A second sample of $\rhod$ events was generated
with a $Q^{-6}$ dependence for the $ep$ reaction, a
flat helicity angular distribution, and an exponentially
falling $p_T^2$ distribution
with a slope of $b$ = 5 GeV$^{-2}$.
The Monte Carlo generator
used the  HERWIG framework~\cite{herwig} and the events were
weighted according to the measured helicity, $p_T^2$ and $y$
distributions.

A third $\rhod$ Monte Carlo generator (RHODI), based on the
model of Forshaw and Ryskin~\cite{forsh},
was used to model the proton dissociative processes with a
$d \sigma(\gamma^* p)/d M_X^2 \propto 1/M_X^{2.5}$ dependence. Different
$M_X$ dependences were obtained by weighting the events. Events were
generated for $M_X^2$ values between 1.2 and 4000 GeV$^2$.
All Monte Carlo events were passed through the
standard ZEUS detector and trigger simulation programs
and through the event reconstruction software.

 The radiative corrections were calculated to be
(10-15)\% for the selection cuts used in the analysis and for the
$Q^2$ and $W$ dependences found in the data. They are taken into account in
the cross sections given below.

\section{ Analysis and cross sections}

\subsection{ Data selection}

\begin{figure}
\epsfysize=18cm
\centerline{\epsffile{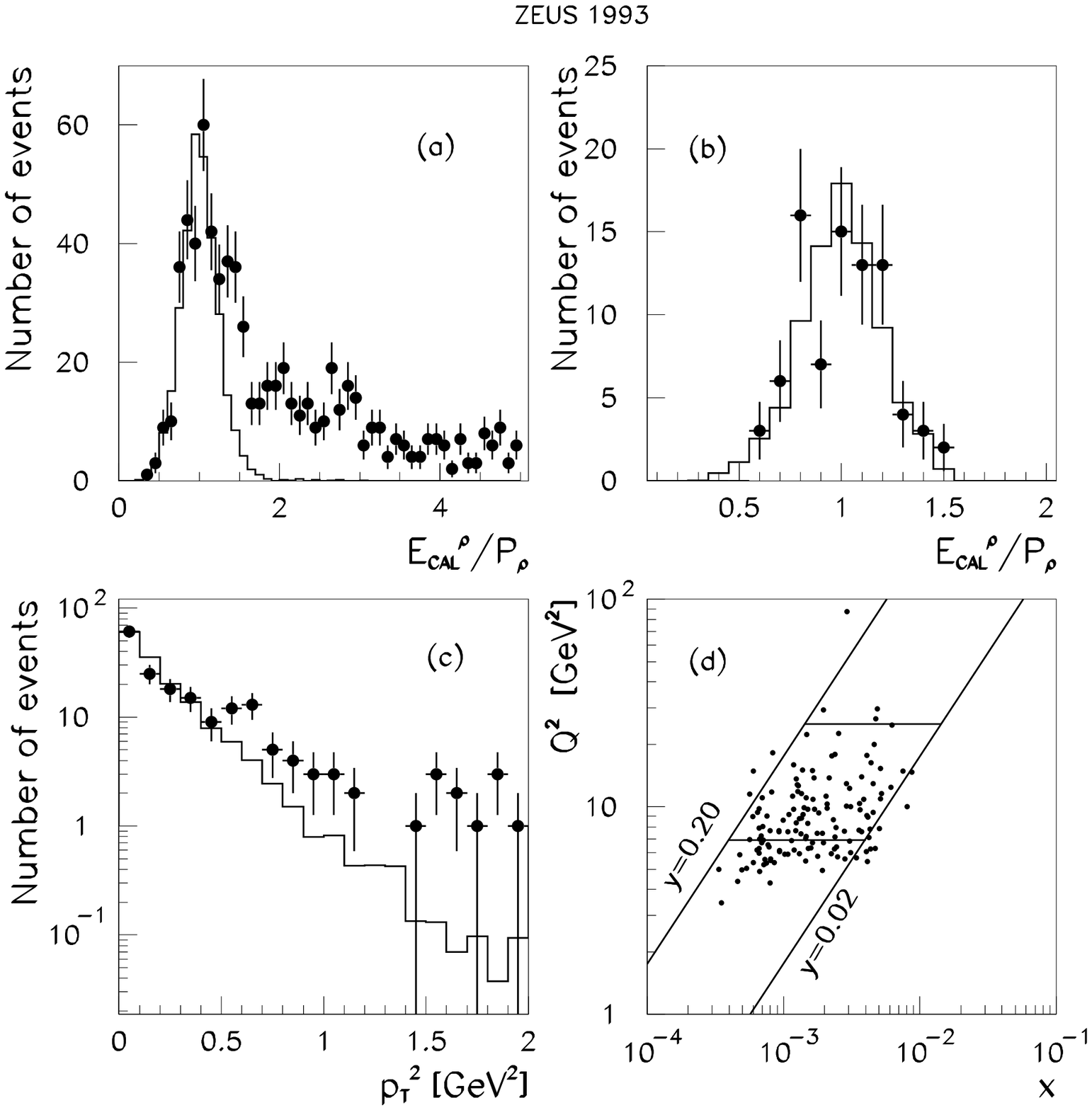}}
\caption{\label{fig1}{
(a) The distribution of $E_{CAL}^{\rho}/P_{\rho}$ for the candidate
events; (b) the same
distribution for the final sample of 82 events;
(c) the $p_T^2$ distribution; and
(d) a scatter plot of $Q^2$ versus $x$ for the selected $\rhod$ events.
These plots are not corrected for detector and
trigger efficiencies and acceptances.
The histograms in (a-c) are obtained from the DIPSI $\rhod$ Monte Carlo
sample after detector and trigger
simulation. In (d) the lines correspond to the region in
$Q^2$ and $y$ selected for this analysis.
}}
\end{figure}

For the selection of exclusive $\rhod$ candidates, the off-line
analysis required:
\begin{itemize}
\item a scattered electron energy, as measured in the calorimeter,
   greater than 5 GeV. The electron identification algorithms used in this
   analysis were optimised to have high efficiency ($>97$\%);
\item  $\delta~=~\sum_i E_i(1 -\rm{cos}\theta_i) ~>~35$ GeV, where
   the sum runs over all calorimeter cells; this cut reduces the
      radiative corrections and photoproduction background;
\item two tracks with opposite charge, both associated
      with the reconstructed vertex; if there was a third
      track at the vertex, it should be from the scattered electron.
      Each of the two tracks was required to have a transverse momentum above
      0.16 GeV and a polar angle between $\rm 25^o$ and $\rm 155^o$; this
      corresponds to the region where the CTD response and systematics are
      well understood;
\item a measured vertex ($Z_{vtx}$), as reconstructed
      from VXD and CTD tracks,  to be in the range $-50 < Z_{vtx} <$ 40 cm;
\item events with a scattered electron whose impact point in the RCAL
      was outside the square of $32 \times 32~$cm$^2$   centered on the
      beam axis or events with an RHES impact point outside the square of
      $26 \times 26~$cm$^2$; this requirement
      controls the determination of the electron scattering angle; and
\item the residual calorimeter energy not associated with the electron
      to be compatible with the $\rhod$ momentum measured in the tracking
      system, $E_{CAL}^{\rho}/P_{\rho}<$ 1.5 (see Fig.~\ref{fig1}a),
      where $E_{CAL}^{\rho}$ is the calorimeter energy excluding that due
      to the scattered electron
      and $P_{\rho}$ is the sum of the absolute values of the momenta of the
      two oppositely charged tracks. This cut suppresses backgrounds with
      additional calorimeter energy unmatched to the tracks and
      events with proton dissociation depositing energy in the calorimeter
      towers around the FCAL beampipe. Also shown in Fig.~\ref{fig1}a
      is the distribution from
      the $\rhod$ DIPSI Monte Carlo events, indicating that only a small
      fraction of the exclusive $\rhod$
      events are removed by this cut. Fig.~\ref{fig1}b
      shows the same distribution
      after the final selection indicating good agreement with the expected
      distribution.
\end{itemize}
A total of 352 events passed these selection requirements.

Possible backgrounds to the exclusive reaction (1) are from $\rhod$ events
with additional undetected particles, from $\phi$ and $\omega$ production
and from proton dissociation events where $M_X$ is small and therefore does
not deposit energy in the calorimeter.
To reduce these backgrounds, two additional cuts were imposed:
\begin{itemize}
\item $0.6 <M_{\pi^+\pi^-}<1.0$ GeV; this selection
     reduces the contamination from $\phi$ and $\omega$ production in
     the low $\pi^+\pi^-$ mass region as well as
     higher mass resonant states and non-exclusive events
     in the high mass region; and
\item $p_T^2 < 0.6$ GeV$^2$; this cut reduces non-exclusive background
      and proton dissociation events. Fig.
      \ref{fig1}c shows the measured, uncorrected
      $p_T^2$ distribution for the selected $\rhod$
       events, indicating a clear excess of events above
      a single exponential for $p_T^2 > $ 0.6 GeV$^2$, consistent with
      the presence of proton dissociation events which, in hadron-hadron
      scattering, have
      a less steep $p_T^2$ distribution~\cite{goul}. (The acceptance is
      relatively flat, rising by about 5\% from $p_T^2$ = 0 to 0.6 GeV$^2$.)
\end{itemize}

Fig.~\ref{fig1}d shows a scatter plot of $Q^2$ versus $x$
for the 140 events which pass the above criteria.
The efficiency
drops sharply at small $Q^2$ (due to the cut on the electron
impact point in RCAL)
and at small and large $y$ (due to the requirements on the
$\pi^+,\pi^-$ tracks).
To remove poorly reconstructed events and to
select a region of phase space where the acceptance is well determined and
relatively constant as a function of the kinematic variables,
two additional kinematic cuts were applied to the data:
\begin{itemize}
\item $7 < Q^2 < 25$ GeV$^2$ and $ 0.02 < y <  0.20$.
\end{itemize}
The final $\rhod$ sample contains 82 events.

\subsection{Background estimates and acceptance corrections}

The DIPSI $\rhod$ Monte Carlo simulated events were used to correct the
data for acceptance and detector resolution. The acceptance
(which includes the geometric acceptance, reconstruction efficiencies,
detector efficiency and resolution, corrections for the offline
analysis cuts and a correction for the $M_{\pi^+\pi^-}$ cut)
in this region of $Q^2$ varies between 40\% and 55\%. The
acceptance is constant at about 47\% as a function of $y$, $p_T^2$ or
$M_{\pi^+\pi^-}$ in the above kinematic region.
The resolutions in the measured kinematic variables, as determined
from the Monte Carlo events, are
6\% for $Q^2$ and 2\% for $y$.

Fig. \ref{fig2}a, which shows the uncorrected $\pi^+\pi^-$ mass distribution
for the events
passing all of the final cuts (except for the $M_{\pi^+\pi^-}$ cut, but with
a $M_{K^+K^-}~>$ 1.05 GeV cut, when the tracks are assigned a kaon mass,
to remove $\phi$ events),
indicates a pure sample of $\rhod$ events.
A $\rhod$ non-relativistic Breit-Wigner form, with a constant background,
is fit to the mass spectrum between 0.6 and 2.0 GeV.
The resulting parameters for the $\rhod$ mass and width are
$774\pm 9$ MeV and $134 \pm 20$ MeV, respectively,
to be compared with the values 769.9 MeV and 151.2 MeV
from the Particle Data Group \cite{pdg}. The fit also includes a flat
background estimate of $(4\pm4)$\% for $M_{\pi^+\pi^-}$
masses between 0.6 and 1.0 GeV,
as determined from comparing the
Monte Carlo $\rhod$ events with the data
for $M_{\pi^+\pi^-}$
masses between 1.0 and 1.5 GeV.
Background contributions from exclusive
$\phi$ and $\omega$ events were estimated to
be at the 1\% level and are included in the above background estimate.

Since the proton was not detected, the proton dissociation background
contribution had to be subtracted. This was done using
the RHODI event generator combined with the detector simulation. The
normalisation
was obtained by requiring that the Monte Carlo generated sample have the
same number of events with energy
between 1 and 20 GeV in the FCAL
as for the data (7 events)
when the constraint that $E_{CAL}^{\rho}/P_{\rho} <$ 1.5 was relaxed to
$(E_{CAL}^{\rho}-E_{FCAL})/P_{\rho} <$ 1.5 and the additional constraint
$\theta_{\pi^{\pm}}> 50^o$ was imposed.
Assuming an $M_X$ dependence of
the form $1/M_X^{2.25}$, as measured by the CDF experiment \cite{cdf} for
$\overline{p}p \rightarrow \overline{p}$ + X,
yielded a contribution of $(22\pm8\pm 15)$\% where the systematic
error was obtained from varying the exponent of $1/M_X$ between 2 and 3.
This is consistent with
an estimate from the excess above the exponential in the $p_T^2$
distribution mentioned above.  In the exclusive $\rhod$ sample under study
here, there are no events with an electron energy in the range
$5<E_e^{\prime}<14$ GeV and so the
photoproduction background is
negligible. No events were found from the unpaired
bunches demonstrating that the beam-gas background is also negligible.
The overall background contamination
was estimated to be $\Delta = (26\pm18)$\%. Unless explicitly
stated otherwise, this background was subtracted as a constant
fraction for the cross sections given below.

\subsection{The $ep$ cross section}

The cross section, measured in the kinematic region defined above, is
obtained from
$ \sigma (e p\rightarrow e \rhod p) =N (1-\Delta)C_1/(C_2 \cdot
A \cdot L_{int}) $,
where $N$ (= 82) is the observed number of events after all cuts
with 0.6 $<M_{\pi\pi}<$ 1.0
GeV, $\Delta$ is the background estimation,
$A$ is the acceptance as discussed above,
$L_{int}$ is the integrated luminosity of 0.55 pb$^{-1}$ and
$C_2$ is the correction for QED radiative effects. These radiative
corrections were
calculated for the exclusive reaction using the $x$ and $Q^2$ dependences
found in this experiment (see section 6.1) and vary between
1.10 (at low $Q^2$ and low $y$) and 1.14 (at high $Q^2$ and high $y$). A
systematic uncertainty of $\pm$0.10 was included on this correction
to account for the uncertainties in the cross section dependences on $x$ and
$Q^2$.
To compare later with results from the NMC experiment, which has determined
exclusive $\rhod$ cross sections integrated over all $p_T^2$~\cite{nmc},
$C_1$ is a 4.5\% correction for the cross section in the
$p_T^2$ range between 0.6 and 1.0 GeV$^2$ based on the slope of the
distribution measured in the present analysis.  The
corrected $ep$ cross section for exclusive $\rhod$ production at $\sqrt{s}$
= 296 GeV is
\begin{displaymath}
   \sigma(e p\rightarrow e \rhod p) =
         0.21 \pm 0.03(stat.) \pm 0.06(syst.) \rm ~nb,
\end{displaymath}
integrated over the ranges $7 < Q^2 <$ 25 GeV$^2$,
$0.02 < y < 0.20$ and $p_T^2 < 1.0$ GeV$^2$, with acceptance corrected
$<Q^2>$ and $<W>$ of 11.0 GeV$^2$ and 78.9 GeV, respectively.

The quoted systematic uncertainty is derived from the following (the
systematic uncertainty for each item is indicated in parentheses):
\begin{itemize}
\item  the cuts used to remove non-exclusive backgrounds were varied and
      independent analyses using differing selection cuts and
  background estimates were compared to the previously described analysis:
  tracks were matched to the calorimeter
  energy deposits and events containing an energy in excess of that
  of the $\rhod$ of more than 0.4, 1 or 2 GeV were discarded;
   events were selected based on the position of
  the electron measured by
  the calorimeter rather than by the RHES
  and the cut on the impact position of the electron in RCAL was varied (10\%);
\item using different trigger configurations (8\%),
\item the cuts on the tracks were varied. The lower cut on the transverse
  momentum was varied between 0.1 and 0.2 GeV and
  different polar angle selections were made. The maximum variation
  occurred for $p_T > 0.2$ GeV (9\%); and
\item events from different Monte Carlo generators
   were used to calculate the acceptance and efficiency (7\%).
\end{itemize}
Adding these in quadrature to those from the uncertainties due to background
subtraction (24\%), luminosity (3.3\%) and
radiative corrections (10\%)
yields 31\% as the overall
systematic uncertainty.

\subsection{The $\gamma^* p$ cross sections}

The $ep$
cross section was converted to a $\gamma^*p$ cross section as follows.
The differential $ep$ cross section for one photon exchange can be expressed
in terms of the transverse and longitudinal virtual photoproduction
cross sections (see \cite{joos}) as:
\begin{displaymath}
\frac{d^2\sigma (ep)}{dxdQ^2} = \frac{\alpha}{2 \pi x Q^2}
\left[\left( 1+(1-y)^2  \right) \cdot
\sigma_T^{\gamma^* p}
(y, Q^2) + 2(1-y)\cdot \sigma_L^{\gamma^*p}(y, Q^2)\right].
\end{displaymath}
The virtual photon-proton cross section can then be written in terms of
the electron-proton differential cross section:
\begin{equation}
\label{sigtot}
\sigma (\gamma^*p \rightarrow \rhod p) =
(\sigma_T^{\gamma^*p}+\epsilon
\sigma_L^{\gamma^*p}) = \frac{1}{\Gamma_T} \frac{d^2\sigma (ep\rightarrow
  e \rho^o p)}{dxdQ^2}
\end{equation}
where $\Gamma_T$, the flux of transverse virtual photons, and $\epsilon$,
the ratio of the longitudinal to transverse virtual photon flux, are given by
\begin{displaymath}
\Gamma_T = \frac{\alpha \left( 1+(1-y)^2 \right) }
{2\pi x Q^2} ~~~{\rm and }~~~
 \epsilon = \frac{2(1-y)}{\left( 1+(1-y)^2 \right) } .
\end{displaymath}
Throughout the kinematic range studied here, $\epsilon$ is in the range
$0.97 < \epsilon < 1.0$.

Using Eq. (\ref{sigtot}),
 $\sigma({\gamma^* p\rightarrow \rhod p}$)
was determined with the flux calculated from the $Q^2$, $x$  and $y$ values
on an event-by-event basis. The 31\% overall systematic uncertainty
on $\sigma(ep)$ applies to every value for
 $\sigma({\gamma^* p\rightarrow \rhod p}$)
and thus becomes an overall normalisation uncertainty.

\section{Results}

\subsection{$Q^2$ and $p_T^2$ distributions}
\begin{figure}
\epsfysize=18cm
\centerline{\epsffile{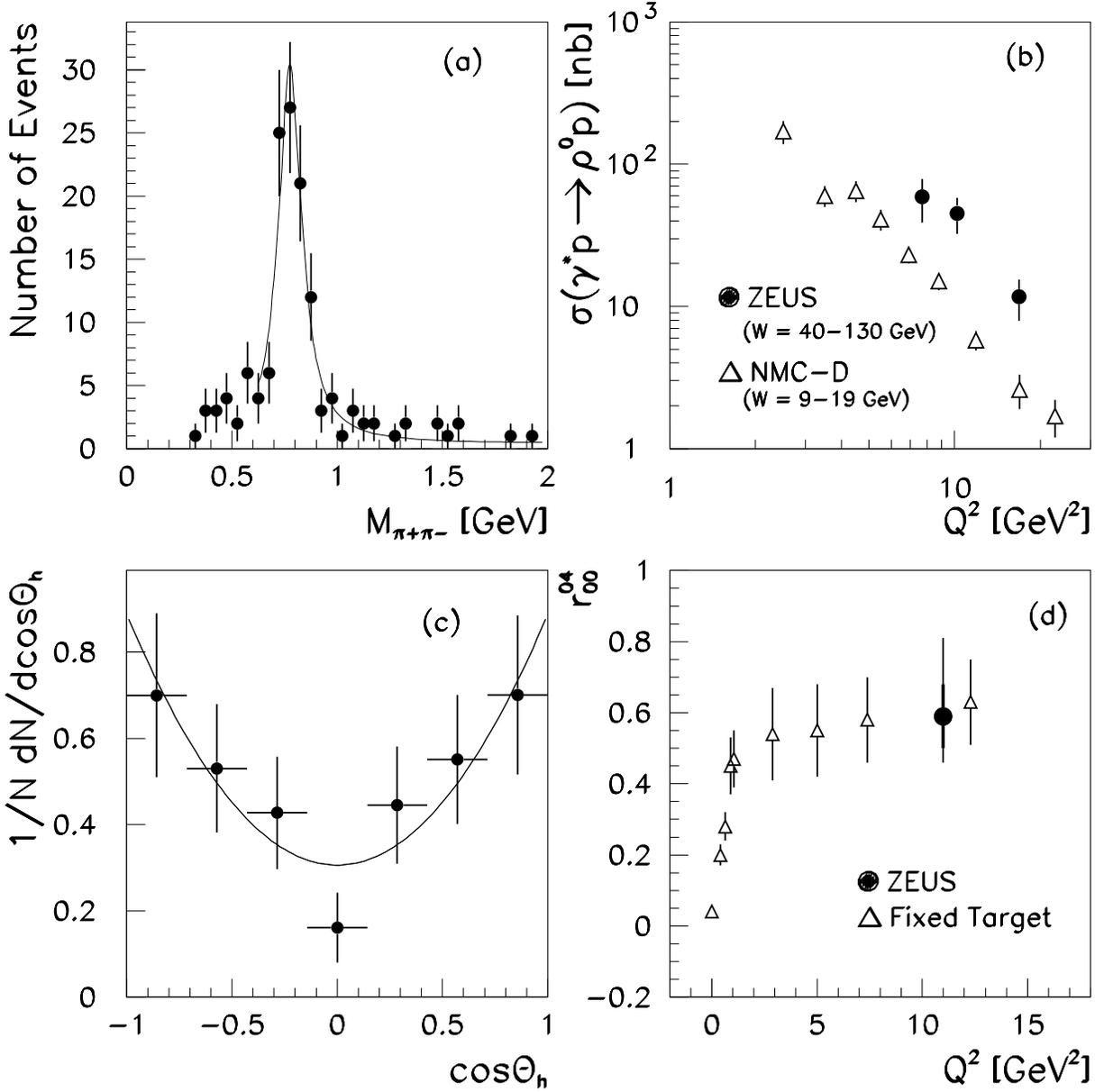}}
\caption{\label{fig2}{
(a) The $\pi^+\pi^-$ invariant mass distribution for the final
   sample of events; the curve is a maximum likelihood fit with a
   non-relativistic Breit-Wigner distribution plus a
   flat (4\%) background (see text for details);
(b) the cross section for $\gamma^* p \rightarrow \rhod p$ as a
   function of $Q^2$ for $0.0014<x<0.004$.
   Also shown are data from the NMC experiment~[7];
   the errors shown are just the statistical errors. The ZEUS (NMC) data
   have an additional 31\% (20\%) normalisation uncertainty
   (not shown);
(c) the cos$\theta_h$ distribution for the decay $\pi^+$, in the s-channel
    helicity system, corrected for acceptance, for
   $\pi^+\pi^-$ pairs in the mass range 0.6-1.0 GeV. The curve is a fit to
   the form of Eq. (10);
(d) the $\rhod$ density matrix element, $r_{00}^{04}$, compared with results
   from fixed target experiments [2,3,7] as a function of $Q^2$. The thick
   error is the statistical error and the thin error is the systematic error
   added in quadrature.
}}
\end{figure}

After correcting for detector acceptance and backgrounds,
the cross sections were obtained
as a function of $Q^2$.
Fig.~\ref{fig2}b displays the $Q^2$ dependence of
the $\gamma^*p \rightarrow \rho^o p$ cross section for events in the $x$
range between 0.0014 and 0.004.  Also displayed in
Fig.~\ref{fig2}b are data from the NMC experiment~\cite{nmc}.
The ZEUS values of the cross sections
are larger than those of the NMC experiment.
However, it should be noted that for this figure as well as for Figs.
2d, 3 and 4 the different experiments have different mixtures of
longitudinal and transverse photon fluxes
($\epsilon$ varies from 0.5-0.8 for the NMC results).
More importantly, the region of $\gamma^* p$ c.m.
energy of the NMC experiment (8-19 GeV) is
lower than that in this experiment (40-130 GeV).

A fit of the form $Q^{-2\alpha}$ to
the distribution of the ZEUS data
in Fig. \ref{fig2}b yields the power of the $Q^2$ dependence.
To study the systematic uncertainty on the value of $\alpha$, a
maximum likelihood analysis was performed with the cross section factorised as:
\begin{equation}
\label{maxlike}
 \sigma(\gamma^* p \rightarrow \rhod p) \sim (Q^2)^{-\alpha} \cdot  x^{-\beta}
  \cdot e^{-b p^2_T}.
\end{equation}
This study, applied to the 82 events in the final data sample in the region
$0.02<y<0.20$, $p_T^2<0.6$ GeV$^2$ and $7<Q^2<25$ GeV$^2$,
yields results similar to those obtained from fitting Fig. \ref{fig2}b.
The best estimate of the $Q^2$ dependence is $ 2\alpha =  4.2 \pm 0.8 (stat.)
 ^{+1.4}_{-0.5}(syst.)$, where the systematic uncertainty comes from
the variation in the value of $\alpha$ obtained from the two different
fitting methods as well as from the variation obtained from the systematic
studies described in section 5.3.

The uncorrected $p_T^2$ distribution was presented in Fig.~\ref{fig1}c and
showed an exponentially falling behaviour, with an excess of events for
$p_T^2 > 0.6$~GeV$^2$, as discussed previously.  After correcting for
detector acceptance and resolution, a fit in the range $0<p_T^2<1.0$
GeV$^2$ of the form:
\begin{equation}
\label{pt}
  \frac{d\sigma}{dp_T^2} = A e^{(-bp_T^2)} + B e^{(- \frac{b}{2}p_T^2)},
\end{equation}
was performed. In Eq.~(\ref{pt})
the contribution from the second term, which was constrained to be
22\% of the number of events for $p_T^2< 0.6 $ GeV$^2$, represents
the proton dissociative background contribution which was assumed to
have a slope half that
of the exclusive reaction~\cite{goul}. The fit yields
a slope parameter
of $b$ = $5.1^{+1.2}_{-0.9} (stat.) \pm 1.0 (syst.)$ GeV$^{-2}$,
which is consistent with that found in the maximum likelihood fit.
The systematic uncertainty comes from the variation in the value
of $b$ obtained from fits without the second term in Eq.~(\ref{pt}),
the maximum likelihood fit and from the systematic studies described
in section 5.3.
This value of $b$ is about half that found in elastic $\rhod$
photoproduction~\cite{spprho} and is in agreement with
the result from the NMC experiment~\cite{nmc}.

\subsection{$\rhod$ decay distribution }

The $\rhod$ s-channel helicity decay angular distribution
$H$(cos$\theta_h,\phi_h,\Phi_h)$ can be used to
determine the $\rhod$ spin state \cite{guenter}, where
$\theta_h$ and $\phi_h$ are the polar and azimuthal angles, respectively,
of the $\pi^+$
in the $\rhod$ c.m. system and
$\Phi_h$ is the azimuthal angle of the $\rhod$ production plane with
respect to the electron scattering plane.
The quantisation axis is defined as the $\rhod$ direction
in the $\rm \gamma^*p$ c.m. system.
Only the  cos$\theta_h$ dependence is presented here.
After integrating over  $\phi_h$ and $\Phi_h$,
the decay angular distribution can be written as:
\begin{equation}
\label{helicity}
    \frac {1}{N}\frac{dN}{d\rm{cos} \it \theta_h} =
   \frac{3}{4}[1-r_{00}^{04}+(3 r_{00}^{04}-1)\rm{cos}^2\it \theta_{h}],
\end{equation}
where the density matrix element $r_{00}^{04}$ represents
the probability that the $\rhod$ was produced
longitudinally polarised by either transversely
or longitudinally polarised virtual photons.

The helicity cos$\theta_h$ distribution (uncorrected for background, since
the dominant contribution to the background is from proton dissociation which
is expected to have the same helicity as the $\rhod p$ final state)
is shown in Fig.~\ref{fig2}c. The curve shown in
the figure is from a maximum likelihood fit to the form of Eq.
(\ref{helicity}) yielding
$r_{00}^{04}=0.6 \pm 0.1 ^{+0.2}_{-0.1}$
at $<Q^2>$ = 11.0 GeV$^2$ and $<W>$ = 78.9 GeV,
where the first uncertainty is statistical,
and the second is derived
from the variations of the result when different ranges in cos$\theta_h$
were used in the fit and when the systematic studies of section 5.3
were used.
 In Fig. \ref{fig2}d this measurement of $r_{00}^{04}$ is compared to
other published data at various values of $Q^2$. These data show
the presence of both transversely and
longitudinally polarised $\rhod$'s at high $Q^2$ (above 2 GeV$^2$).
If SCHC is assumed,
an estimate of $R$, the ratio of longitudinal to transverse cross sections,
for $\rhod$ production is obtained~\cite{joos}:
\begin{displaymath}
     R = \frac{\sigma_L}{\sigma_T} = \frac{1}{\epsilon}\cdot \frac{r_{00}^{04}}
     {1-r_{00}^{04}} = 1.5 ^{+2.8}_{-0.6}
\end{displaymath}
(where the statistical and systematic uncertainties in $r_{00}^{04}$ have
been added in quadrature).
This may be compared with the value of $R = 2.0\pm 0.3$ at $<Q^2>$ = 6 GeV$^2$
and $<W> \sim $ 14 GeV from the NMC experiment~\cite{nmc}.

\subsection{The $W$ and $x$ dependences of the $\gamma^*p \rightarrow
\rhod p$ cross section}

\begin{figure}
\epsfysize=18cm
\centerline{\epsffile{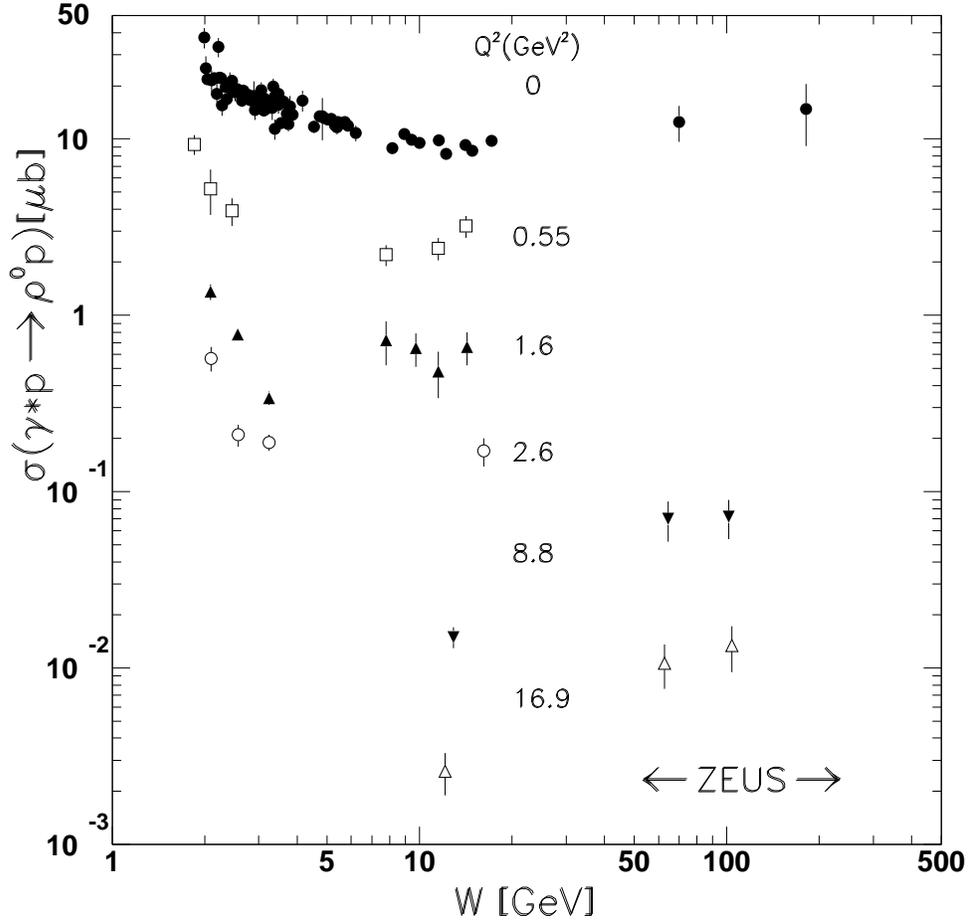}}
\caption{\label{fig3}{
The $\gamma^* p \rightarrow \rhod p$ cross section as a function of
$W$, the $\gamma^*p$ centre of mass energy, for several
values of $Q^2$. The low energy data ($W<$ 20 GeV) come from fixed target
experiments [2,4,5,7,8]. The high energy data  ($W>$ 50 GeV) come from the
ZEUS experiment [10] and the present analysis. The ZEUS data at $Q^2$ = 8.8
and 16.9 GeV$^2$ have an additional
31\% systematic normalisation uncertainty (not shown); the data from Refs.
[2], [4] and [7] have additional 10\%, 25\%, and 20\%
normalisation uncertainties, respectively.
}}
\end{figure}

Fig. \ref{fig3} shows a compilation [2,4,5,7-9] of photoproduction and selected
leptoproduction exclusive $\rhod$ cross sections
as a function of both $Q^2$ and $W$.
In this figure the cross sections obtained in this analysis are shown as a
function of $W$ at $Q^2$ = 8.8 and 16.9 GeV$^2$.
The cross sections at different $W$ values (and slightly different $<Q^2>$)
were scaled to
$Q^2$ = 8.8 and 16.9 GeV$^2$ using the measured $Q^{-4.2}$ dependence in order
to compare with the NMC cross sections\footnote{Since the EMC and NMC
data cover approximately the same kinematic region, the more recent NMC
data\cite{nmc} have been chosen to make comparisons.} from deuterium
at the same values of $Q^2$.
At high energies, $W>$ 50 GeV, data exist only at $Q^2$ = 0, 8.8 and 16.9
GeV$^2$. The real ($Q^2$ = 0) $\gamma p$
 `elastic' cross section~\cite{spprho} shows only a slow rise, consistent
with that seen in the photon-proton total cross section.
At small $Q^2 ~(< $ 2.6 GeV$^2$), the data first
decrease with increasing $W$ followed by a slow increase.
No high energy data are yet available to see how the increase
develops.
At higher $Q^2$, the cross sections rise strongly with increasing
$W$.  At $Q^2$ = 8.8 GeV$^2$
and $W \sim 100$ GeV, the cross section is about a factor of
six larger than at $W$ =  12.9 GeV~\cite{nmc}.
This strong increase in the $\gamma^* p \rightarrow \rhod p$ cross
section is in contrast to that expected from the Donnachie and Landshoff
model~\cite{dl1,dl2}.

\begin{figure}
\epsfysize=18cm
\vskip 4cm
\centerline{\epsffile{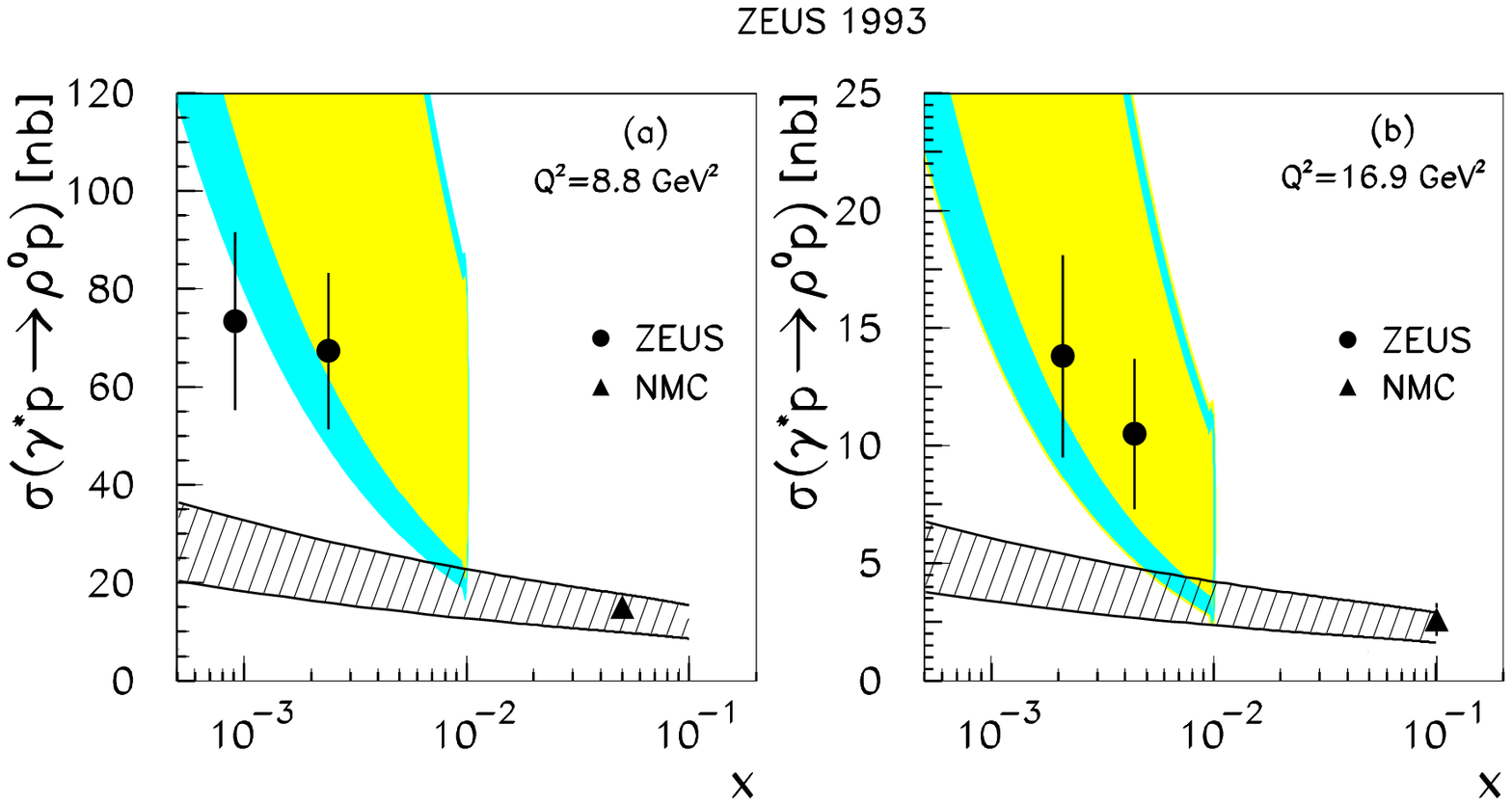}}
\vskip -6cm
\caption{\label{fig4}{
(a) The cross section, $\sigma (\gamma^* p \rightarrow \rhod p)$, as
   a function of $x$ at a value of $Q^2$ = 8.8 GeV$^2$.
(b) A similar plot for data at $Q^2$ = 16.9 GeV$^2$.
    The errors shown are only the statistical errors. In addition, there
   is a 31\% systematic uncertainty
   which is not shown but applies to the overall normalisation.
   Also shown is the NMC result
   (which has an additional 20\% normalisation uncertainty~[7]).
   The shaded area corresponds to the predictions of
   Eqs. (3) and (11) for $x<0.01$. The range of the predictions shown by
   the light shaded area is a result
   of the experimental uncertainty of the gluon
   distribution~[18]. The dark shaded area includes the uncertainties
   on $b$ and $R$ added in quadrature to that of the gluon.
   In addition, there is a 50\% uncertainty in the predicted
   cross section from the choice of the $Q^2$ scale; furthermore, the value
   of $x$ in Eq. (3) is only defined to within a factor of 2.
   The hatched area   displays the cross section expected from the
   soft pomeron model [11]. The range comes from the uncertainty in the
   measured value of $b$.
}}
\end{figure}

To compare with the QCD calculations of Brodsky et al. and Donnachie and
Landshoff, the ZEUS and NMC cross sections are shown as a
function of $x$ at $Q^2$ = 8.8 and 16.9 GeV$^2$ in Figs. \ref{fig4}a, b.
The total cross sections are predicted from the
pQCD calculations of Brodsky et al.~\cite{brod}
using the
longitudinal contribution to the differential cross section at $t$ = 0,
(see Eq. (\ref{brodsky})):
\begin{equation}
\label{sigbrod}
    \sigma (x, Q^2) = \int ^1_0 \frac{d\sigma(x,Q^2,p_T^2)} {dp_T^2} dp_T^2 =
       \frac{(1 - e^{-b})}{b} \cdot
\left. \frac{d(\sigma_T+\sigma_L)}{dt} \right|_{t=0}
     =       \frac{(1 - e^{-b})} {b} \cdot (\frac{1}{R}+\epsilon) \cdot
\left. \frac{d\sigma_L}{dt}\right|_{t=0}
\end{equation}
where $b$ is the slope of the $p_T^2$ distribution. The
measured values of $\epsilon$ = 0.98, $R$ = 1.5 and $b$ = 5.1
 GeV$^{-2}$ were used to calculate $\sigma(x,Q^2)$.
For the gluon momentum density, $xg(x,Q^2)$, the form
$xg \sim x^\delta \cdot (1-x)^\eta$ was assumed.
The values of $\delta$ and $\eta$ were
allowed to vary within the ranges ($-0.25$ to $-0.39$) and (6.44 to 4.81)
respectively. These ranges were determined from
the leading order gluon density extracted~\cite{zgluon} from the scaling
violations of the ZEUS F$_2$ measurements~\cite{zf2}.
The uncertainty in Eq.~(\ref{sigbrod}) arising from the $1\sigma$ range in the
gluon density\cite{zgluon} is shown as the light shaded region in Fig.
\ref{fig4}. The uncertainty in the prediction arising from measurement
uncertainties of $R$ and $b$ when added in quadrature to that of
the gluon distribution yields the larger dark shaded area.
The shaded areas in Fig. \ref{fig4} are restricted
to $x < 0.01$ where Eq. (\ref{brodsky}) is valid.
The range in the predicted cross sections at a given $x$ is dominated by
the uncertainty on $\delta$. Since the calculation is made in DLLA, the
value of $Q^2$ at which the gluon density and $\alpha_s$ are evaluated is not
defined to better than a factor of two. Varying the $Q^2$ scale for
$\alpha_s(Q^2) \cdot xg(x,Q^2)$ from $Q^2$/2
to $2Q^2$ changes the prediction by about 50\%. The $x$ scale in the
calculation can also
range from $x/2$ to $2x$ so that the curves can be shifted left or right to
reflect this uncertainty.
At the present level of precision of the measurement and theory, the data are
consistent with the pQCD calculation of Brodsky et al.
The hatched region shows the range of the predictions of Donnachie and
Landshoff based on soft pomeron exchange \cite{dl1}
with $\sigma = \frac{1}{b} \left. \frac{d\sigma}{dt}\right|_{t=0}$. The
range of the hatched area comes from the uncertainty in the measured
value of $b$.
The data do not agree with these expectations, being typically a factor of
three above the predictions.

\section{Conclusions }

Exclusive $\rhod$ production has been studied in deep inelastic
electron-proton scattering at large $Q^2$ (7 - 25 GeV$^2$) in the
$\gamma^*p$ centre of mass energy ($W$) range from 40 to 130 GeV.
Cross sections are given for both the
$ep$ and $\gamma^*p$ processes. The cross section for the $\gamma^*p$ process
exhibits a $Q^{-(4.2 \pm 0.8 ^{+1.4}_{-0.5})}$ dependence.
The $\gamma^* p \rightarrow \rhod p$ cross section at these large $Q^2$ values
shows a strong increase with $W$ at HERA
energies over the lower energy NMC data, in contrast to the
behaviour of the elastic photoproduction cross section. Both longitudinally
and transversely polarised $\rhod$'s are produced. The $Q^2$ dependence,
the polarisation and  the slope of the $p_T^2$ distribution are
consistent with those observed at lower energies. However, the cross
sections are significantly larger.
The Donnachie and Landshoff prediction for soft pomeron exchange
underestimates the measured cross
sections while the data are consistent with the perturbative
QCD calculation of Brodsky et al. given the present knowledge of the
gluon momentum density in the proton.

\vspace{1cm}
\noindent {\Large\bf Acknowledgements}

\vspace{1cm}

The strong support and encouragement of the DESY Directorate
is greatly appreciated.
The experiment was made possible by the inventiveness and the diligent
efforts of the HERA machine group who continued to run HERA most
efficiently during 1993. We also acknowledge the many informative
discussions we have had with S. Brodsky, J. Cudell, J. Forshaw, L. Frankfurt,
J. Gunion, P. Landshoff, A. Mueller, M. Ryskin, A. Sandacz and M. Strikman.


\newpage
\parskip 0mm

\newpage
\parskip 0mm

\newpage
\parskip 0mm

\newpage
\parskip 0mm


\begin{thebibliography}{1}
\bibitem{bauer} T. H. Bauer et al., Rev. Mod. Phys. 50 (1978) 261.
\bibitem{joos} P. Joos et al., Nucl. Phys. B113 (1976) 53.
\bibitem{slac} C. del Papa et al., Phys. Rev. D19 (1979) 1303.
\bibitem{lame} D. G. Cassel et al., Phys. Rev. D24 (1981) 2787.
\bibitem{chio} W. D. Shambroom et al., Phys. Rev. D26 (1982) 1.
\bibitem{emc} EMC Collab., J. J. Aubert et al., Phys. Lett. 161B (1985) 203;
         \newline          J. Ashman et al., Z. Phys. C39 (1988) 169.
\bibitem{nmc} NMC Collab., P. Amaudruz et al., Z. Phys. C54 (1992) 239;
    \newline  M. Arneodo et al., Nucl. Phys. B429 (1994) 503.
\bibitem{sbt} J. Ballam et al., Phys. Rev. Lett. 24 (1970) 960.
\bibitem{rhot} ``Total Cross-Sections for Reactions of High Energy Particles'',
     Landolt-B\"ornstein, New Series, Vol. 12b, editor H. Schopper (1987).
\bibitem{spprho} ZEUS Collab., M. Derrick et al., ``Measurement of the
  elastic $\rhod$ photoproduction cross section at HERA", paper ICHEP94 Ref.
  0688, submitted to the 27th International Conference on High Energy Physics,
  Glasgow (1994); \newline
  ZEUS Collab., M. Derrick et al., Z. Phys. C63 (1994) 391.
\bibitem{dl1} A. Donnachie and P. V. Landshoff, Phys. Lett. B185 (1987) 403.
\bibitem{dl2} A. Donnachie and P. V. Landshoff, Nucl. Phys. B311 (1989) 509;
     Phys. Lett. B348 (1995) 213.
\bibitem{cudell} J. R. Cudell, Nucl. Phys. B336 (1990) 1.
\bibitem{ryskin} M. G. Ryskin, Z. Phys. C57 (1993) 89; and private
   communication.
\bibitem{ginz} I. F. Ginzburg et al., ``Semihard quasidiffractive production
     of neutral mesons by off shell photons", preprint, submitted to
     Nucl. Phys. B (1994).
\bibitem{nnz} J. Nemchik et al., Phys. Lett. B341 (1994) 228.
\bibitem{brod} S. J. Brodsky et al., Phys. Rev. D50 (1994) 3134.
\bibitem{zgluon}  ZEUS Collab., M. Derrick et al., Phys. Lett. B345 (1995) 576.
\bibitem{zf2}  ZEUS Collab., M. Derrick et al., Z. Phys. C65 (1995) 379.
\bibitem{vxd} C. Alvisi et al., Nucl. Instr. and Meth. A305 (1991) 30.
\bibitem{ctd} C. B. Brooks et al., Nucl. Instr. and Meth. A283 (1989)
      477;  \newline B. Foster et al., Nucl. Instr. and Meth. A338 (1994) 254.
\bibitem{hes} ZEUS Collab., The ZEUS Detector Status Report 1993.
\bibitem{lumi} J. Andruszk\'ow et al., DESY 92--066 (1992).
\bibitem{wsmith} W. H. Smith et al., Nucl. Instr. and Meth. A355 (1995) 278.
\bibitem{arneo}  M. Arneodo, L. Lamberti and M. G. Ryskin,
            to be submitted to Comp. Phys. Comm. (1995).
\bibitem{herwig} G. Marchesini et al.,  Comp. Phys.  Comm. 67 (1992) 465.
\bibitem{forsh} J. R. Forshaw and M. G. Ryskin, DESY 94-162 and private
communication.
\bibitem{goul} K. Goulianos, Phys. Rep. 101 (1983) 169.
\bibitem{pdg} Review of Particle Properties, Particle Data Group, Phys. Rev.
    D50 (1994) 1664.
\bibitem{cdf} CDF Collab., F. Abe et al., Phys. Rev. D50 (1994) 5535.
\bibitem{guenter} K. Schilling, P. Seyboth and G. Wolf, Nucl. Phys. B15
(1970) 397; \newline K. Schilling and  G. Wolf, Nucl. Phys. B61 (1973) 381.


\end{thebibliography}
\end{document}